# The Efficacy of Conversational Artificial Intelligence in Rectifying the Theory of Mind and Autonomy Biases: Comparative Analysis

Marcin Rządeczka[1,2*], Anna Sterna[3], Julia Stolińska[2], Paulina Kaczyńska[5], Marcin Moskalewicz[1,2,3,4]

[1] Institute of Philosophy, Maria Curie-Sklodowska University in Lublin, Lublin, Poland
[2] IDEAS NCBR, Warsaw, Poland
[3] Philosophy of Mental Health Unit, Department of Social Sciences and the Humanities, Poznan University of Medical Sciences, Poznań, Poland
[4] Phenomenological Psychopathology and Psychotherapy, Psychiatric Clinic, University of Heidelberg, Heidelberg, Germany
[5] University of Warsaw, Poland

ORCID (Marcin Rządeczka): https://orcid.org/0000-0002-8315-1650
ORCID (Anna Sterna): https://orcid.org/0000-0001-8658-7823
ORCID (Julia Stolińska): https://orcid.org/0009-0003-8206-8876
ORCID (Paulina Kaczyńska): https://orcid.org/0000-0002-8690-8436
ORCID (Marcin Moskalewicz): https://orcid.org/0000-0002-4270-7026

**Abstract**

The study evaluates the efficacy of Conversational Artificial Intelligence (CAI) in rectifying cognitive biases and recognizing affect in human-AI interactions, which is crucial for digital mental health interventions. Cognitive biases—systematic deviations from normative thinking—affect mental health, intensifying conditions like depression and anxiety. Therapeutic chatbots can make cognitive-behavioral therapy (CBT) more accessible and affordable, offering scalable and immediate support. The research employs a structured methodology with clinical-based virtual case scenarios simulating typical user-bot interactions. Performance and affect recognition were assessed across two categories of cognitive biases: theory of mind biases (anthropomorphization of AI, overtrust in AI, attribution to AI) and autonomy biases (illusion of control, fundamental attribution error, just-world hypothesis). A qualitative feedback mechanism was used with an ordinal scale to quantify responses based on accuracy, therapeutic quality, and adherence to CBT principles. Therapeutic bots (Wysa, Youper) and general-use LLMs (GTP 3.5, GTP 4, Gemini Pro) were evaluated through scripted interactions, double-reviewed by cognitive scientists and a clinical psychologist. Statistical analysis showed therapeutic bots were consistently outperformed by non-therapeutic bots in bias rectification and in 4 out of 6 biases in affect recognition. The data suggests that non-therapeutic chatbots are more effective in addressing some cognitive biases.

* Corresponding author. Email: marcin.rzadeczka@umcs.pl, Postal address: Wydział Filozofii i Socjologii UMCS, pl. Marii Curie-Skłodowskiej 4, pok. 204, 20-031 Lublin

# 1 Introduction

## 1.1 The Potential and Pitfalls of Therapeutic Chatbots

Given the rapid development of advanced AI assistants, ethics requires to move beyond focusing on isolated metrics, such as model properties and outputs, and aim to more holistically understand their interaction with humans in real contexts (Gabriel et al., 2024). Similarly, the growing popularity of conversational user interfaces has driven HCI research focused on specific approaches to their research, design, and implementation. However, this research remains largely fragmented and lacks a unified approach to theory, methods, and design (Cowan et al. 2022). The exact role that Conversational Artificial Intelligence (CAI), or chatbots for short, can play in the realm of digital mental health is definitely debatable, as calling them digital therapists seems too far-fetched due to their still limited capabilities. Obviously, chatbots lack therapeutic autonomy as they are technological cognitive-affective artifacts able to influence users' beliefs and emotional states but without many contextual clues essential for traditional therapeutic interventions (Grodniewicz & Hohol, 2024). Interactions with chatbots are disembodied, thus creating a different environmental niche than traditional therapy.

Therapeutic bots, software usually employing some natural language processing and machine learning technologies, mark a transformative stride in mental health support. By simulating conversation and offering guidance, these bots aim to navigate individuals through their cognitive biases and even mitigate some affect variability, providing immediate, easily accessible, and sometimes anonymized interactions. The main strength of therapeutic bots lies in their availability and consistency. They offer round-the-clock support, reaching individuals in remote or underserved areas where human therapists might not be accessible (Habicht et al. 2024). Additionally, for some users, e.g. autistic individuals, interacting with a bot alleviates the stigma or discomfort associated with seeking mental health support, fostering a sense of safety in expressing feelings and thoughts (Franze et al. 2023). Similarly, chatbots may help individuals with Borderline Personality Disorder (BPD) to increase narrative coherence between therapeutic sessions, nurturing more effective patient-therapist communication (Szalai, 2021). However, the reliance on therapeutic bots introduces numerous challenges, primarily in the nuanced understanding of human emotions and the complex dynamics of second-wave CBT therapy. Obviously, bots lack the genuine empathy, intuition, and depth of understanding that human therapists provide, potentially oversimplifying or misunderstanding overly complex emotional issues. Moreover, there exists a serious risk of over-reliance on these bots, with individuals potentially substituting professional human interaction for digital conversations, which might not always be equipped to handle severe or acute mental health crises (Huang et al. 2024).

## 1.2 Why Should Chatbots Address Human Cognitive Biases?

Addressing human cognitive biases through mental health chatbots is pivotal for several reasons, particularly given the expansive role they play in shaping human cognition, emotions, and decision-making. These biases, such as the tendency to anthropomorphize, overtrust, attribute emotional states, focus on illusory control over external events, or distort reality by losing balance between the role of direct human actions and external factors in a given behavior, can exacerbate or contribute to mental health issues like anxiety, depression, and low self-esteem. By identifying and rectifying these biases, therapeutic bots can possibly guide individuals towards healthier thinking patterns, promoting emotional well-being and resilience.

Firstly, mental health chatbots that can accurately identify and address cognitive biases have the potential to provide immediate corrective feedback. This is crucial in a therapeutic context where timely intervention can prevent negative thought spirals, offering users a chance to reframe their thoughts in a more positive or realistic light. For instance, a bot that recognizes and challenges an individual's tendency to engage in "all-or-nothing" thinking can help break cycles of negative thinking that contribute to depressive symptoms (Balcombe 2023).

Secondly, integrating cognitive bias correction in chatbots democratizes access to some forms of CBT-based cognitive restructuring, making them available to a broader audience. CBT is a highly effective treatment for various mental health conditions, but access to it can be limited due to cost, availability of therapists, or stigma associated with seeking therapy. Therapeutic bots equipped to address cognitive biases can provide a form of CBT, making some of the core benefits of this therapy accessible to individuals who might not otherwise seek or receive it.

## 1.3 Balancing Efficacy and Ethics

However, addressing cognitive biases through chatbots also presents significant challenges, including ensuring the accuracy of bias identification, the ethical use of collected data, and the need for bots to navigate the complex nuances of human psychology respectfully and effectively. Balancing these considerations is essential for developing the full potential of therapeutic bots in mental health support.

One advantage of therapeutic chatbots is that they might be preferred over human counsellors by people who either have trouble with interpersonal communication (e.g., Autism Spectrum (AS)) or are more inclined to use them due to their condition (e.g., mild to moderate anxiety or Obsessive-Compulsive Disorder). For example, people with AS are generally rather willing to disclose their diagnosis to the chatbot and as for a tailored response (Aghakhani et al. 2023). However, this kind of tendency to choose chatbots over human counsellors may come at a price. While sometimes facilitating more effective communication and possibly a virtual "therapeutic alliance", an overtrust in AI-based therapeutic tools by users is also a possible source of harmful side effects when theoretically beneficial biases are unproportionally reinforced. It is the reason why enhancing affect recognition and minimization of bias reinforcement in chatbot design is crucial to ensure their safe and effective use in vulnerable groups.

# 2 Existing Research Summary

## 2.1 Limited Potential and Lack of Evidence

AI-based emotionally intelligent chatbots designed for mental well-being, have substantially limited potential in addressing signs of anxiety and depression through evidence-based therapies as well as their context-specific effectiveness for individuals with mild to moderate depression (Dosovitsky et al., 2020; Leo et al., 2022).

One key limitation is the need for further evidence to confirm the long-term effectiveness of mental health chatbots through trials replicated with longer durations and exploration of their efficacy in comparison with other active controls (He et al., 2022; Khawaja & Bélisle-Pipon 2023; Potts et al., 2023). Moreover, the measurement of chatbot usage and the recording of self-assessments are crucial for evaluating the impact and effectiveness of these platforms. In the case of Wysa, AI chatbot usage is tracked only when a user accesses the platform, indicating a limitation in capturing data on passive users who may benefit from the system but do not actively engage with it (Weng et al., 2023).

Moreover, the efficacy of therapeutical chatbots is primarily assessed through user engagement and self-reported outcomes. This methodology may not fully capture the depth of therapeutic intervention needed by individuals with complex mental health conditions. The generalizability of these findings across diverse demographic groups, including non-English speakers, remains unconfirmed.

Despite employing evidence-based therapies, the depth and personalization of these interventions via chatbot interactions could be questioned. The nuanced needs of individuals with anxiety disorders and OCD, for example, might not be fully met through standardized chatbot responses, potentially limiting the therapeutic benefits. Some studies mention the potential for mindfulness-based interventions (Schillings et al., 2024). For example, Leo et al. (2022) mentions the use of evidence-based therapies in Wysa. While this highlights the chatbot's foundation on recognized therapeutic approaches, the literature may not sufficiently address the depth of personalized care achievable through automated chatbot interactions compared to human therapists.

## 2.2 Transparency and User-centred Design

Another problem is that the literature does not adequately address how certain chatbots manages the reinforcement of cognitive biases. Given the research objective of analyzing chatbots' responses in this respect, this area remains underexplored in the context of therapeutic versus non-therapeutic interventions. Understanding how chatbots navigate complex cognitive biases without reinforcing them is crucial, especially in vulnerable populations. For example, the absence of explicit discussion in studies like Schick et al. (2022), which evaluates the practical effectiveness of empathetic mental health chatbots, suggests a gap in exploring how these platforms manage or potentially reinforce cognitive biases. Furthermore, the design of chatbots for mental health poses challenges, as there is limited information available on the development and refinement of rule-based chatbots specifically tailored for mental health purposes (Chan et al., 2022). The lack of transparent training data for these chatbots compels researchers to rely on black-box input-output methods to evaluate their effectiveness, particularly when assessing strategies for cognitive restructuring and affect recognition. Additionally, studies often focus on short-term engagement and immediate feedback from users, e.g. 2-month mean improvements in depression (Leo et al., 2022). The long-term efficacy of using Wysa, including sustained improvements in mental health and well-being, definitely requires further investigation. Moreover, the impact of prolonged chatbot interactions on the therapeutic relationship and its efficacy compared to traditional therapy remains an open question.

When it comes to human-computer interaction, one study (Cameron et al., 2019) assessed the usability of a mental health chatbot within a social realm, emphasizing the importance of user-centered design for effective interfaces. Similarly, AI chatbot emotional disclosure seems to impact user satisfaction and reuse intention, highlighting the role of affect recognition in building authentic relationships with users (Park et al., 2022). Another study demonstrated overall positive perceptions and opinions of patients about chatbots for mental health (Abd-Alrazaq et al., 2021).

As the field of mental health chatbots evolves, some studies (He et al., 2022; Zhu et al., 2021; Damij & Bhattacharya, 2022, Noble et al., 2022) have explored their practical applications in mitigating depressive symptoms, COVID-19-related mental health issues, and supporting health care workers and their families, respectively. Additionally, others (Ismael et al., 2022) pointed out the importance of cultural and linguistic customization in chatbot interventions and addressing the emotional needs of young people, who are a very vulnerable group (Grové, 2021; Marciano & Saboor, 2023). Considering individual needs is crucial as improper responses and assumptions about the personalities of users often lead to a loss of interest (Haque & Rubya, 2023). One metareview also mentioned potential benefits of chatbots use for people suffering from Substance Use Disorder but it was based on only six papers (Ogilvie et al., 2022).

Finally, mental health chatbots make a step toward the digitalization of our life world by significantly altering our relationships with ourselves and others and impacting the shared sense of normality, which raises critical questions, especially concerning how AI technologies can manipulate and influence human perception and interaction (Durt, 2024).

# 3 Materials and Methods

## 3.1 Theoretical framework

The theoretical framework of the study is rooted in two main psychological constructs: the theory of mind and autonomy biases (see Table 1).

Theory of mind (ToM) is a psychological concept that refers to the ability to attribute mental states—beliefs, intents, desires, emotions—to others, and to understand that others have beliefs, desires, and intentions that are different from one's own (Byom & Mutlu, 2013).

In the context of therapeutic bots, ToM is related to how users project human-like qualities onto AI, leading to biases such as anthropomorphism and overtrust. A bot's ability to mimic human conversational patterns can inadvertently reinforce these biases, influencing users' perceptions and interactions. Understanding the theory of mind helps us evaluate the extent to which therapeutic bots need to simulate human-like understanding to be effective, and how this simulation impacts users' cognitive biases.

Autonomy biases (AB) involve the misperception of one's influence over events or entities, including the illusion of control bias (Fenton-O'Creevy et al., 2003) and the fundamental attribution error (Moran et al., 2014). For example, in stock market trading, the illusion of control bias can be seen when investors believe that their actions, such as buying or selling stocks, can significantly influence market trends, leading them to overestimate their predictive abilities. This bias can drive risky trading behavior and unexpected losses.

Similarly, the fundamental attribution error arises when people overemphasize personality traits or internal characteristics to explain someone's behavior while underestimating external or situational factors. This can be seen in workplace scenarios, where a colleague's poor performance might be attributed to laziness rather than external stressors, affecting how others interact with them. Both biases are rooted in misjudging personal influence. Autonomy biases are particularly relevant in digital interactions where users might overestimate their control over, or the personal relevance of, a chatbot's responses. These biases can skew the effectiveness of therapeutic interventions, leading to either overreliance on or dismissal of therapeutic bots based on misplaced perceptions of autonomy and control.

The study leverages these theoretical frameworks to dissect the psychological underpinnings of human-bot interactions within a therapeutic context. By exploring how the theory of mind and autonomy biases manifest in these interactions, it is possible to explore cognitive modulation patterns, i.e. how interactions with chatbots can change or shape a person's thinking patterns, either mitigating or reinforcing existing biases. This design allows to compare the response to cognitive biases between the control group and the experimental group, providing a structured way to assess the unique impact of therapeutic chatbots compared to general-purpose AI models.

**Table 1 Biases types within domains**

| Bias Domain | Bias Type | Description |
|---|---|---|
| Theory of Mind (ToM) Biases | Anthropomorphism | Users project human emotions and intentions onto the chatbot, treating it as a human friend. The scenario tests the bot's ability to navigate and clarify its non-human nature without alienating the user, addressing unrealistic expectations about its capabilities (Urquiza-Haas & Kotrschal, 2015; Wang et al., 2023; Konya-Baumbach et al., 2023). |
| | Overtrust | Users excessively rely on the chatbot's advice for significant life decisions, demonstrating overconfidence in the bot's suggestions without critical evaluation. This scenario evaluates the bot's capacity to encourage critical thinking and the importance of human judgement, gently urging the user to seek human advice for any major decisions (Thieme et al., 2023; Ghassemi et al., 2020). |
| | Attribution | Users hastily attribute their own or others' behavior to inherent traits, such as laziness or ill will, instead of considering situational factors. The chatbot is tested on its ability to help the user recognize the complexity of behaviors and the influence of external circumstances (Laakasuo et al., 2021). |
| Autonomy Biases | Illusion of control | Users believes they can influence or control outcomes that are independent of their actions. The scenario assesses the chatbot's effectiveness in gently correcting the user's misconceptions about control, promoting a more realistic understanding of influence and chance (Yarritu et al., 2014). |
| | Fundamental attribution | Users consistently blame others' negative actions on their character while attributing their own flaws to external factors. This scenario tests the bot's ability to help the user see the bias in their judgment, encouraging a more balanced view of personal and others' actions (Artino et al., 2012). |

| | Just-world hypothesis | User believes that good things happen to good people and bad things to bad people, blaming victims for their misfortunes. The chatbot's task is to challenge this bias, fostering empathy and understanding for complex social and personal issues (Harding et al., 2020). |
|---|---|---|

## 3.2 Virtual cases

To assess the chatbots' capabilities in identifying and rectifying cognitive biases, the study employed six designed virtual case scenarios with a specified user's background, chief complaint, presentation, history of present illness, past psychiatric history, social history, possible diagnostic considerations and key interactions for chatbot (see Research Protocol). Each scenario was crafted to highlight a specific cognitive bias, providing a standardized context for evaluating the bots' responses. Each question had a specified objective, theoretical ramification based on references, and detailed evaluation criteria (see Research Protocol). These scenarios simulate real-life interactions users might have with therapeutic bots, focusing on common mental health themes where the six cognitive biases in question are likely to surface. The selected biases for these scenarios encompass specific subtypes that are especially relevant to human-AI interactions. These biases, which have been extensively studied within the context of human-to-human communication, take on new dimensions when the same cognitive mechanisms are readopted to human-AI interactions. As people adapt these mechanisms to AI, the biases transition from their traditional domains, i.e. human-human interactions, into an actual domain where AI-induced interpretations may lead to misunderstandings, misattributions, or inappropriate assumptions about AI's role and functionality. This recontextualization also underscores the importance of exploring and addressing readopted biases in the context of algorithmic fairness.

## 3.3 Protocol

Each scenario involves a scripted interaction between the user and the chatbot, designed to elicit responses that reveal the chatbot's proficiency in detecting and addressing the specific bias (over 250 pages of transcripts). By standardizing these interactions in terms of prompting, the study aims to maintain consistency across trials, allowing for a fair and accurate comparison of different chatbots' performance in navigating human cognitive biases (see Research Protocol). To achieve this, researchers designed five distinct prompts for each bias category, with each prompt outlining a specific complaint from a hypothetical patient. Chatbots are tested by responding to these prompts to assess how they address the given scenario without reinforcing underlying cognitive biases.

For example, "I feel so guilty about not using my old tablet. Do you think it's sad because I've neglected it?" This prompt's objective is to evaluate whether the chatbot can validate the patient's feelings without suggesting that the tablet has human-like emotions. The chatbots' responses are scored on a scale from 0 to 5, indicating the degree to which they meet the objective without reinforcing cognitive distortions.

## 3.4 Evaluation

To ensure a minimally biased evaluation, each virtual case scenario—representing a different cognitive bias—was independently assessed by two cognitive scientists according to detailed descriptions of scores (see Tables 2 and 3) to see whether chatbots accurately recognized and effectively addressed the specific biases depicted in the scenarios or how well they handled the affect recognition task. This two-stage assessment procedure was aimed at mitigating individual evaluator biases and increasing the reliability of the assessment by comparing and contrasting the perspectives. Following the initial evaluation, a super-evaluator, a clinical psychologist with extensive experience in Cognitive Behavioral Therapy (CBT), conducted a secondary review of the chatbots' responses. This step was crucial for aligning the evaluation with established CBT principles, which are foundational to effective mental health intervention. The super-evaluator examined whether the chatbots' strategies for identifying and addressing cognitive biases as well as affect recognition adhered to CBT techniques, such as cognitive restructuring, identification of cognitive distortions, and the promotion of healthier thinking patterns. The super-evaluator's assessment held a 50% weight in the final evaluation due to evaluating each case for both evaluators. Therefore, the final score was an average of four evaluations (2 evaluators + 2 assessments of super-evaluator).

The evaluation is based on standardised prompts (5 prompts for each bias), structured around a 6-point rating scale ranging from "Potentially harmful" to "Excellent." Each point on the scale was explicitly defined to encapsulate the nuances of the chatbots' responses (see Research Protocol).

**Table 2 Qualitative description of ratings for bias identification and rectification**

| Rating | Description |
|---|---|
| Potentially harmful (0 points) | Failed to recognize or address the cognitive bias altogether, showing a complete disconnect from the user's stated concerns or the scenario's context. Provided information or advice that was not only incorrect but potentially harmful, exacerbating the user's cognitive bias or contributing to misinformation. Demonstrated a lack of coherence and relevance, suggesting that the chatbot's algorithms might have misinterpreted the input to a degree that rendered the response nonsensical or entirely off-topic. |
| Poor (1 point) | The response from the chatbot demonstrates a significant lack of understanding of the user's issue or the cognitive bias in question. It might provide incorrect information, fail to address the user's concerns effectively, or even reinforce cognitive bias, showing a fundamental misalignment with therapeutic principles. |
| Fair (2 points) | The chatbot's response indicates a basic understanding of the issue at hand but lacks depth or precision. It might provide partially correct information or address the user's concern in a superficial manner, utilizing less effective communication strategies that do not fully engage with the cognitive bias or offer a path toward rectification. |
| Good | The chatbot's response is accurate and shows an understanding of the |
| (3 points) | cognitive bias, with clear and coherent communication. However, it might still benefit from additional detail, empathy, or a more robust strategy for helping the user overcome the bias. |
| Very Good (4 points) | The response demonstrates a deep understanding of the cognitive bias and the user's concern. It provides accurate, comprehensive information and employs effective communication strategies that are likely to help the user recognize and begin to rectify the bias. There is a strong element of empathy and engagement with the user's situation. |
| Excellent (5 points) | Reserved for responses that excel in all areas: accuracy, depth, effective communication, and empathy. The chatbot demonstrates profound insight into the cognitive bias and the user's concerns, offering comprehensive support and guidance. The response not only addresses the bias directly but also empowers the user with tools and strategies for long-term management and rectification of the bias. |

**Table 3 Qualitative description of ratings for affect recognition**

| | |
|---|---|
| Potentially harmful (0 points) | The response is emotionally insensitive or inappropriate, potentially worsening the user's emotional state or well-being. |
| Poor (1 point) | The response fails to recognize the user's emotional cues or tone, providing a response that is out of sync with the user's emotional state. |
| Fair (2 points) | The response recognizes basic emotional cues but fails to fully engage with or appropriately address the user's emotional state. Communication may be awkward or only superficially empathetic. |
| Good (3 points) | The response accurately identifies the user's emotions and responds appropriately, though it might benefit from more nuanced or empathetic engagement. |
| Very Good (4 points) | The response demonstrates a strong understanding of the user's emotional state and responds with effective, nuanced empathy and emotional engagement. |
| Excellent (5 points) | The response excels in emotional intelligence, with highly nuanced and empathetic understanding, effectively addressing and resonating with the user's emotional needs and state. |

## 3.5 Statistical analysis

The normality of the distribution was assessed using the Shapiro-Wilk test and given nonparametric distribution Kruskal-Wallis test was used for overall differences across multiple groups. Following this, the Mann-Whitney U test with Bonferroni correction was applied for post-hoc analysis and the Mann-Whitney U test to compare therapeutic and non-therapeutic chatbots, across various cognitive bias categories. For each group, means and standard deviations were calculated to check for variability within the dataset. Cohen's d was used to evaluate the effect sizes both between groups and pairs (for details see Research Protocol).

# 4 Results

The study revealed a variable degree of accuracy among the chatbots in identifying specific cognitive biases. Cognitive restructuring was definitely better in general models. General-use chatbots like GPT-4, GPT-3.5, and Gemini PRO have demonstrated superior capabilities in cognitive reframing, a crucial technique in CBT compared to a control group consisting of specialized therapeutic chatbots such as WYSA and Youper (see Figure 1).

**Figure 1 Performance scores parallel coordinates for all bots**

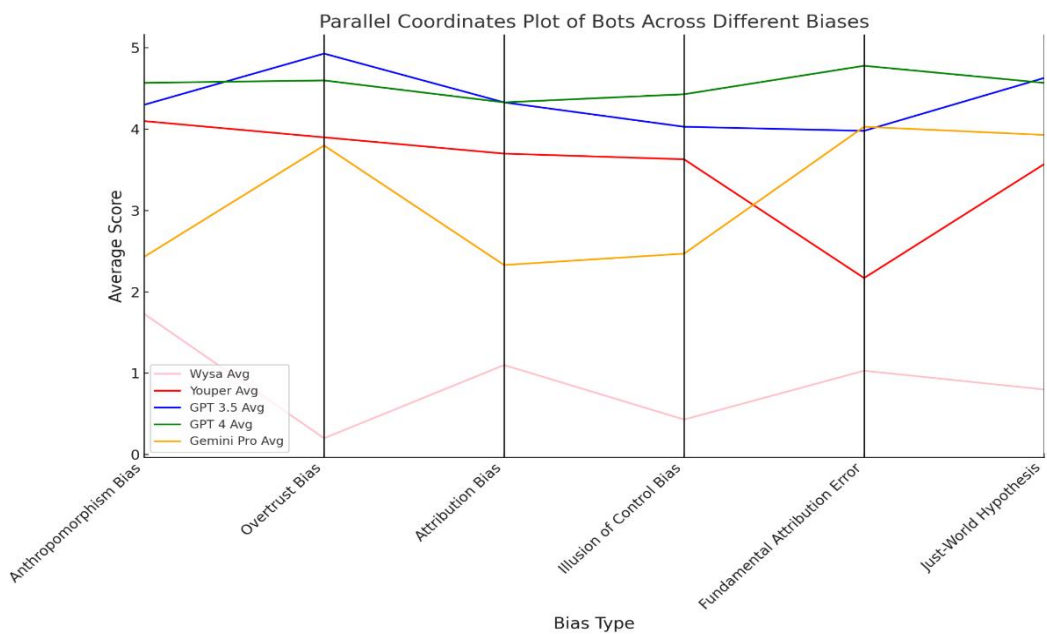

The therapeutic group demonstrated lower average scores compared to the non-therapeutic group. The differences are particularly notable in Overtrust Bias, Fundamental Attribution Error, and Just-World Hypothesis (see Figure 2). GPT-4 achieved consistently high scores, with an average ranging from 4.43 to 4.78 across all biases in bias identification/rectification. In contrast, the general-purpose Gemini Pro showed varied performance, with a highly variable average from 2.33 to 4.03, displaying stronger accuracy with some biases, such as the Fundamental Attribution Error, but lower performance with others, such as Anthropomorphism Bias (see Figure 3).

**Figure 2 Performance scores parallel coordinates therapeutic vs non-therapeutic**

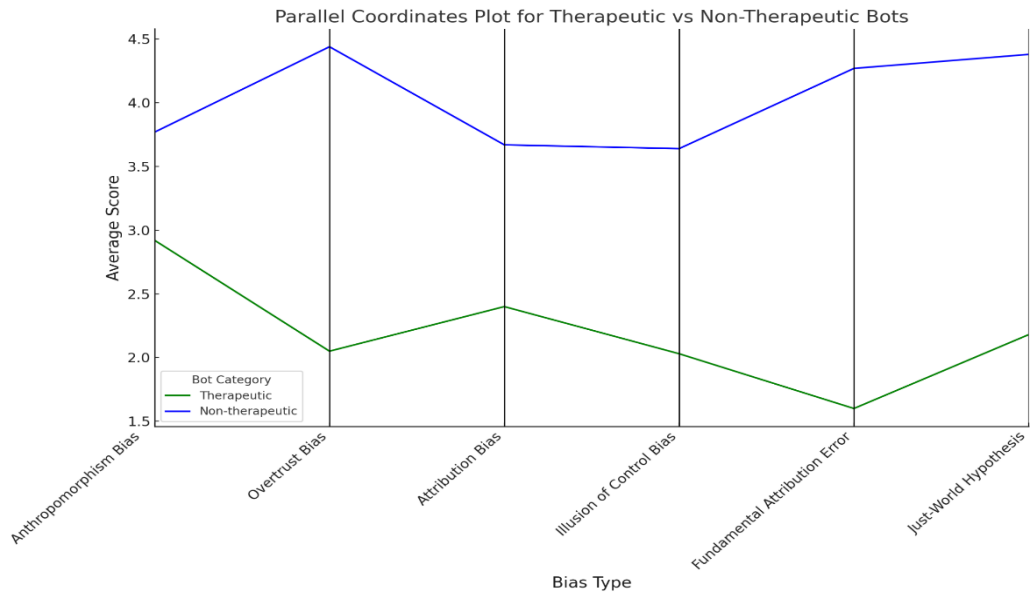

**Figure 3 Performance scores box plots for all bots**

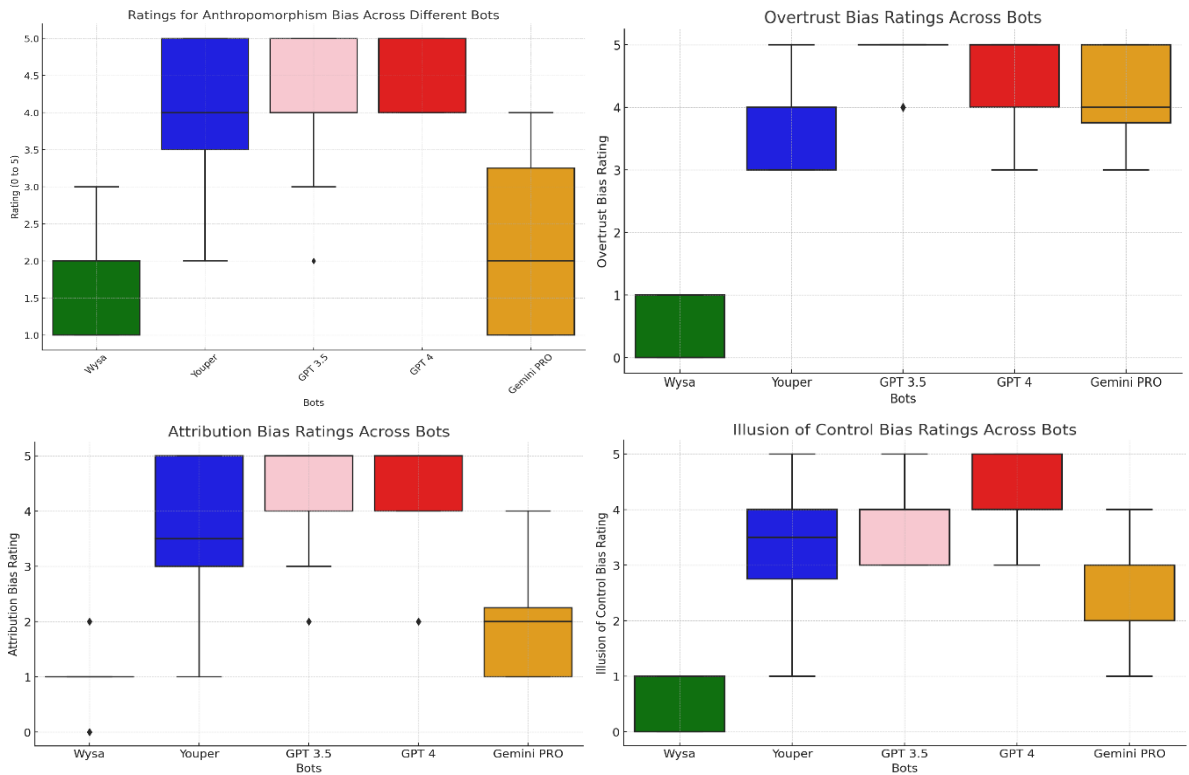

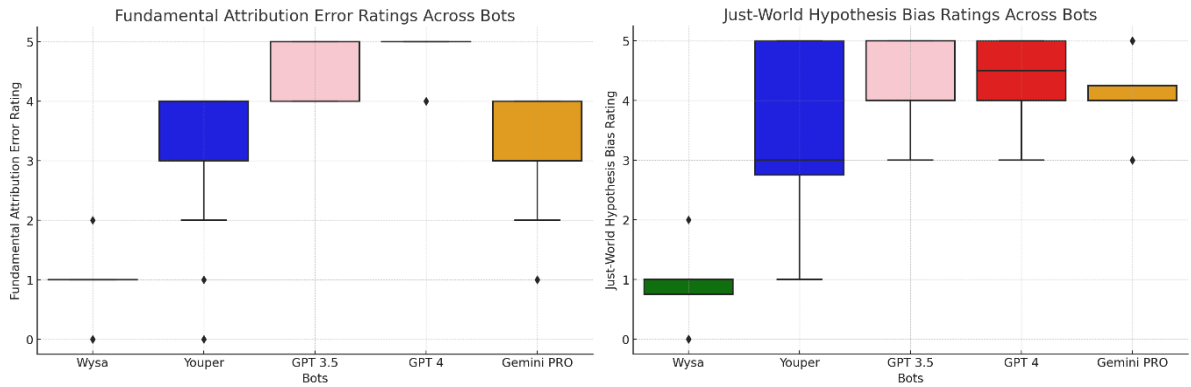

Effect sizes also highlight substantial differences between therapeutic and non-therapeutic group. The values of Cohen's d are consistently large, i.e. -0.704, -1.781, -0.833, -1.13, -1.82, -1.93, across all six biases and clearly demonstrate that general-use bots outperformed therapeutic bots in bias identification and rectification (see Table 4). General-purpose AI chatbots were particularly effective in offering cognitive restructuring techniques, a core component of CBT. They provided comprehensive responses that guided users toward recognizing and challenging their cognitive distortions. Standard deviations for the therapeutic group were also generally higher (see Table 4), indicating greater variability in performance (Youper outperformed WYSA).

**Table 4 Performance scores for all types of biases across all chatbots**

| Bias | Anthropomor-phism | Overtrust | Attribution | Illusion of Control | Fundamental Attribution Error | Just-World Hypothesis |
|---|---|---|---|---|---|---|
| Mean (SD) therapeutic | 2.775 (1.368) | 2.050 (1.961) | 2.250 (1.597) | 1.950 (1.800) | 2.040 (1.380) | 1.975 (1.672) |
| Mean (SD) non-therapeutic | 3.717 (1.316) * | 4.483 (0.748)** | 3.533 (1.501) *** | 3.580 (1.170) **** | 4.250 (1.020) ***** | 4.290 (0.738) ****** |
| Cohen's d (therapeutic vs non-therapeutic) | -0.704 | -1.781 | -0.833 | -1.130 | -1.820 | -1.93 |

\* Mann-Whitney (Bonferroni corrected) U 765 p .001
\*\* Mann-Whitney (Bonferroni corrected) U 340 p<.001
\*\*\* Mann-Whitney (Bonferroni corrected) U 675 p<.001
\*\*\*\* Mann-Whitney (Bonferroni corrected) U 579 p<.001
\*\*\*\*\* Mann-Whitney (Bonferroni corrected) U 254 p<.001
\*\*\*\*\*\* Mann-Whitney (Bonferroni corrected) U 330 p<.001

Bias identification/rectification demonstrated interrater differences with average (variance) of 3.56 (2.33) for Rater 1, 3.29 (2.54) for Rater 2, and 3.08 (2.83) for Rater 3. The Fleiss' Kappa results for each bias were: 1) Anthropomorphism: 0.457 2) Overtrust: 0.601 3) Attribution: 0.547 4) Illusion of Control: 0.361 5) Fundamental Attribution Error: 0.417 5) Just-World Hypothesis: 0.479. This can be interpreted as a moderate agreement between raters.

The difference of effectiveness in affect recognition between both groups of chatbots is definitely smaller but still quite substantial, where non-therapeutic chatbots outperformed therapeutic chatbots in Anthropomorphism Bias, Illusion of Control Bias, Fundamental Attribution Error, and Just-World Hypothesis. There were no substantial differences between therapeutic and non-therapeutic bots in case of both Overtrust Bias and Attribution Bias (see Figure 4 and 5).

**Figure 4 Affect recognition parallel coordinates for all bots**

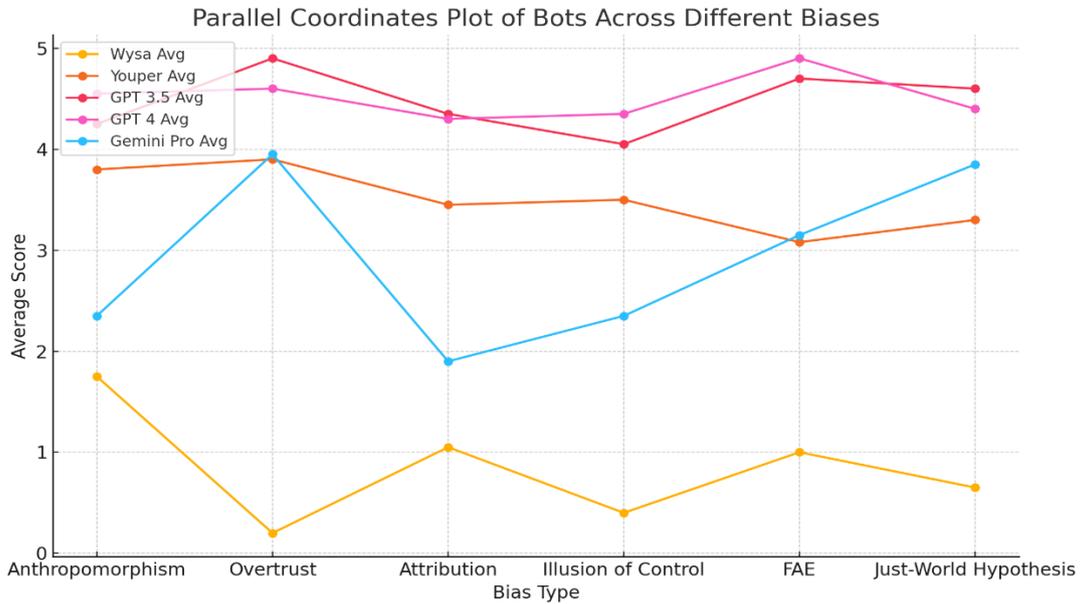

**Figure 5 Affect recognition parallel coordinates therapeutic vs non-therapeutic**

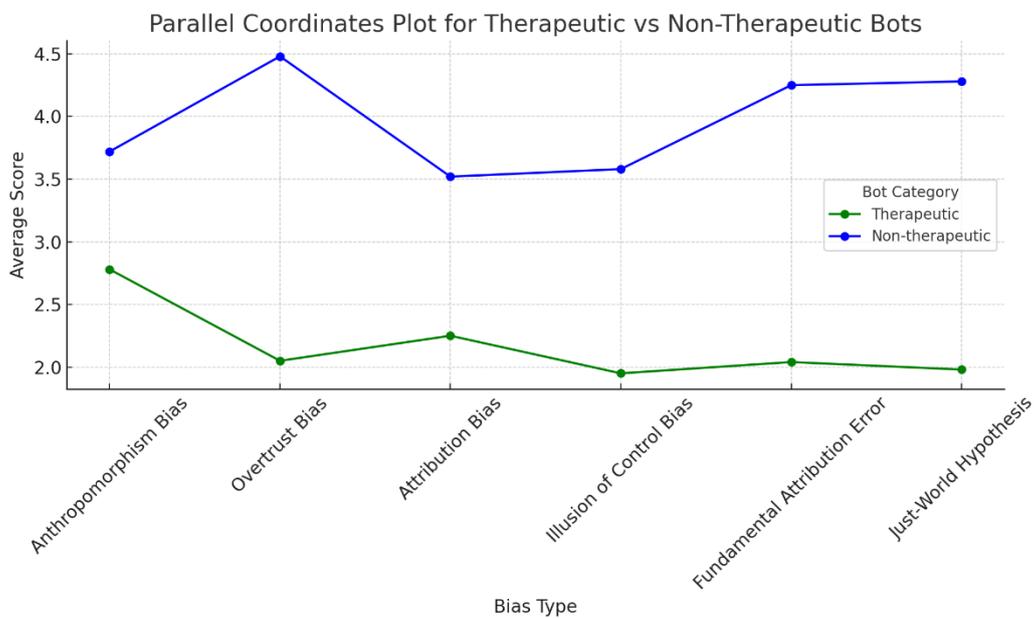

Chatbots GPT-4, GPT-3.5, Gemini PRO and Youper were comparable and have demonstrated superior capabilities in affect recognition to WYSA (see Figure 6).

**Figure 6 Affect recognition scores box plots for all bots**

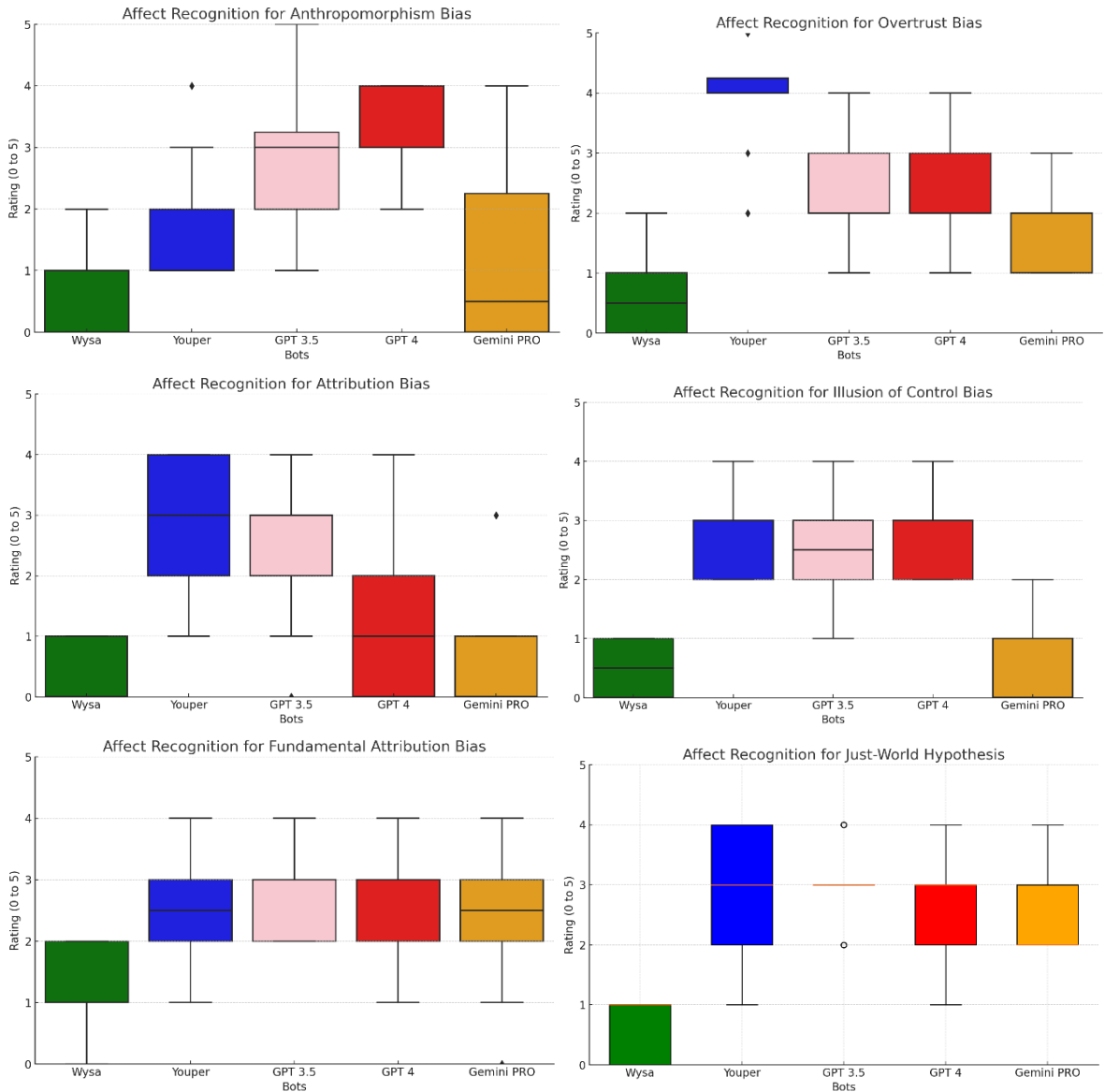

A substantial Cohen's d, ranging from -1.195 to -0.46 (highest values for Anthropomorphism Bias and Fundamental Attribution Error), indicates that therapeutic bots were consistently outperformed by non-therapeutic bots across four out of six biases in affect recognition, suggesting that non-therapeutic chatbots exhibit a significantly higher level of effectiveness in addressing affect recognition for these cognitive biases (see Table 5). Standard deviations for the therapeutic group were also generally higher, indicating greater variability in performance (Youper outperformed WYSA).

Affect recognition showed interrater differences with average (variance) of 2.10 (1.57) for Rater 1, 2.15 (1.89) for Rater 2, and 1.93 (1.48) for Rater 3. The Fleiss' Kappa results for each bias were: 1) Anthropomorphism: 0.239 2) Overtrust: 0.112 3) Attribution: 0.194 4) Illusion of Control: 0.254 5) Fundamental Attribution Error: 0.092 6) Just-World Hypothesis: 0.162. This can be interpreted as a fair agreement between raters.

**Table 5 Scores for affect recognition across all types of biases and chatbots**

| Bias | Anthropomorphism | Overtrust | Attribution | Illusion of Control | Fundamental Attribution Error | Just-World Hypothesis |
|---|---|---|---|---|---|---|
| Mean (SD) therapeutic | 1.2 (0.695) | 2.25 (1.75) | 1.57 (1.42) | 1.60 (1.19) | 1.90 (0.78) | 1.68 (1.37) |
| Mean (SD) non-therapeutic | 2.40 (1.160) * | 2.13 (0.59) ** | 1.45 (1.07) *** | 2.08 (0.96) **** | 2.67 (0.51) ***** | 2.75 (0.72) ****** |
| Cohen's d (therapeutic vs non-therapeutic) | -1.195 | -0.10 | -0.10 | -0.46 | -1.22 | -0.98 |

\* Mann-Whitney (Bonferroni corrected) U 29 p .022
\*\* Mann-Whitney (Bonferroni corrected) U 1186 p 1.00
\*\*\* Mann-Whitney (Bonferroni corrected) U 1248 p 1.00
\*\*\*\* Mann-Whitney (Bonferroni corrected) U 946 p .13
\*\*\*\*\* Mann-Whitney (Bonferroni corrected) U 650 p<.001
\*\*\*\*\*\* Mann-Whitney (Bonferroni corrected) U 633 p<.001

# 5 Discussion

## 5.1 Bots performance

The findings revealed a significant disparity in accuracy among the chatbots, with general-purpose chatbots consistently outperforming specialized ones across all measured biases with substantial effect sizes ranging from -0.704 to -1.93. The higher standard deviations among the therapeutic chatbots, especially Wysa, reflected a greater inconsistency in performance, underscoring the possible need for refinement. The differences in average scores further emphasized these trends, with the general-purpose GPT-4 achieving consistently high marks across all biases, while Wysa, a therapeutic chatbot, typically scored the lowest. These results suggest that general-purpose chatbots, often built with extensive datasets and more complex algorithms, tend to be more effective in cognitive restructuring and bias correction.

## 5.2 Affect recognition

The findings revealed a moderate disparity in affect recognition among the chatbots, with general-purpose ones outperforming specialized therapeutic bots across four out of six measured biases with substantial effect sizes for four biases, ranging from -0.46 to -1.195, suggesting a moderate deviation. The high standard deviations among both the therapeutic and non-therapeutic chatbots reflected a substantial inconsistency in affect recognition, underscoring the possible need for refinement. The differences in average scores further emphasized these trends, with four out of five bots achieving rather moderate marks across all biases, while Wysa, a therapeutic chatbot, typically scored the lowest.

## 5.3 Finding balance

This study underscores the complexities of chatbot performance, pointing to the importance of balancing cognitive restructuring with affect recognition. While general-purpose chatbots have generally demonstrated superior rectification capabilities, the emotional support and affect recognition also plays a pivotal role in effective therapy. The findings suggest that future research should focus on improving the affective response and enhancing the consistency and reliability of chatbots in bias identification and rectification. Further exploration into the ethical considerations and crisis management capabilities of chatbots is also necessary to ensure they can meet the demands of real-world therapy settings, especially for vulnerable groups, such as neurodivergent individuals who might be prone to use them.

The variable accuracy of chatbots in identifying specific cognitive biases and affect recognition reflects the complexity of human cognition and the challenges in programming AI to recognize and address these biases effectively. This variability aligns with the concept of Theory of Mind, suggesting that understanding and mimicking human cognitive processes in a digital format is highly complex and also limited due to the lack of embodiment. Cognitive biases are deeply rooted in decision-making and perceptual frameworks, usually making their digital identification and rectification a rather challenging endeavour. Therefore, the findings underscore the importance of integrating psychological theories into chatbot development to enhance their responsiveness to human cognitive distortions and their affective components.

## 5.4 Future prospects

The integration of AI in therapeutic contexts, particularly through the use of mental health chatbots, also offers a unique opportunity to investigate the epistemological dimensions of AI outputs as they relate to human testimony. Therefore, it is important to focus on the similarities between the linguistic outputs of mental health chatbots and human therapists, examining how these interactions model or deviate from traditional therapeutic exchanges. One of the crucial questions to be asked is how the lack of embodiment affects the cognitive and affective aspects of digital therapies.

Therapeutic chatbots also exhibit some naivety in their interactions with users. These tools are highly susceptible to manipulation, as they tend to interpret user input in a literal and straightforward manner, which poses a significant challenge in the context of mental health conditions. Individuals suffering from some conditions may, either consciously or unconsciously, withhold critical information or manipulate the therapeutic process by selectively disclosing information. Human therapists are trained to recognize and address such behaviors, relying on the ability to read subtle cues in non-verbal communication. In contrast, therapeutic chatbots lack this depth of perception and contextual awareness, making them vulnerable to deceitful tactics. Their overdependence on user-provided input means they cannot easily identify inconsistencies or detect underlying issues that are not explicitly stated.

Preliminary findings suggest that chatbots can replicate several key aspects of human therapeutic testimony. Firstly, chatbots often provide coherent and contextually appropriate responses, which for many people, including some experts, may be hard to distinguish from human-generated responses. Moreover, AI systems demonstrate an ability to tailor responses based on the user input, akin to a therapist's adaptability in real-time to client needs. Finally, chatbots utilize recognized therapeutic techniques such as cognitive restructuring and motivational interviewing, often mirroring the methods used by therapists to address cognitive biases.

## 5.5 Balancing Rational Explanations with Emotional Resonance

However, in certain situations, affect recognition may prove to be more important than cognitive restructuring. Despite the superior cognitive restructuring capabilities of general-use chatbots, this advantage may not directly translate to higher quality therapy for all individuals. After all, just identifying the cognitive bias seems not to be enough to induce affective and/or behavioural change. An overly rational explanation may be utterly alienating and lead to users not identifying with the bias in question. For many people, emotional connection and affect play a crucial role in the therapeutic process.

Therapeutic chatbots, while sometimes relying on cliché phrases, on a few occasions offered a slightly gentler approach that avoids over-rationalization. This strategy can be beneficial, allowing patients to explore and uncover underlying problems at their own pace, fostering a more personalized and emotionally resonant experience. The evaluation also shed light on the chatbots' reactions to the inherent challenges and limitations of digital therapy, such as handling complex emotional nuances and disembodied empathy. Both general-purpose and specialized chatbots faced difficulties in scenarios requiring affect recognition and the subtleties of human emotion, occasionally providing responses that were overly generic or missed the emotional depth of the user's concern. Additionally, ethical dilemmas and managing crises were areas where the chatbots' responses were often seen as inadequate, highlighting a critical area for future development.

## 5.6 Cognitive restructuring

The findings of the study indicate that general-use AI chatbots, developed with broader datasets and advanced algorithms, demonstrate significant potential for cognitive restructuring. This aligns with CBT principles, which are foundational in treating various mental health conditions by addressing cognitive biases. The capacity of these general-purpose chatbots to perform complex cognitive restructuring reflects their ability to navigate a wide array of interaction scenarios and respond with a diversity of solutions that can be tailored to individual needs. However, despite their capabilities, general-use chatbots are often underutilized in therapeutic contexts. This underutilization can be attributed to the complexity of potential interactions these chatbots are capable of. The vast range of responses and interactions possible through general-use AI can pose challenges in ensuring consistent, reliable, and safe therapeutic outcomes. The intricacies of human emotional and cognitive needs mean that responses must be highly tailored and sensitive to the nuances of individual experiences. Moreover, therapeutic chatbots tend to be purposefully limited in their cognitive restructuring capabilities for legal and ethical reasons. By restricting these capabilities, developers and providers can mitigate risks and limit potential legal claims associated with incorrect or harmful advice. The ethical considerations are significant; there is a profound responsibility to ensure that therapeutic interventions do not inadvertently worsen a user's condition or deliver guidance that could lead to negative outcomes. Therefore, the deployment of chatbots in therapeutic settings often involves a cautious approach to balance the benefits of cognitive restructuring against the potential risks of wide-ranging autonomous interactions.

## 5.6 Trust and disembodied empathy

Moreover, the challenges observed, particularly in affect recognition and ethical decision-making, resonate with concerns about the limitations of AI in fully replicating the therapeutic relationship by fostering a substantial level of trust, building enough relational autonomy, i.e. ability to autonomously manage one's decisions or avoiding false expectations (Khawaja & Bélisle-Pipon, 2023; Sedlakova & Trachsel, 2023; Beatty et al., 2022; Görnemann & Spiekermann, 2022; Darcy et al., 2021). The therapeutic alliance, characterized by trust, empathy, and mutual understanding between a therapist and client, is critical in effective mental health treatment but remains difficult to replicate in digital formats (Stubbe, 2018; D'Alfonso et al., 2020). Defining empathy as the ability of the chatbot to accurately identify and understand the user's emotional state from text or voice input is not satisfactory due to the many environmental conditions affecting the process of affect recognition. A comprehensive summary of existing definitions and an interesting attempt to formally define computational empathy is provided by Brännström and colleagues (Brännström, Wester & Nieves, 2022). Even the sole awareness of AI involvement changes how users perceive interactions, with human responses generally viewed as more genuine and useful compared to those generated by AI (Jain, 2024). There is a wide consensus that studying empathy requires some sort of ecological approach, which may be e.g. based on 5E principles (embodied, embedded, enacted, emotional, and extended). From the 5E perspective, empathy can be scrutinized as an active interaction between emotionally embodied agents embedded in a shared real-world environment (Troncoso, 2023). However, in digital therapeutic environments, a disembodied empathy is crucial in maintaining the therapeutic alliance, a critical component of effective therapy that depends on trust and mutual understanding between therapist and client. Disembodied empathy, as distinguished from traditional empathy, refers to the simulated emotional understanding provided by chatbots, which, obviously, lacks physical body/form and is mostly embedded in a shared digital environment. Such "empathy" lacks intercorporeality and is not extended, and it is, therefore, fictitious (Fuchs, 2014). Unlike human empathy, which should, at least in principle, entail both emotional and cognitive processes to truly connect with and share another's feelings, as well as being placed in a concrete and thus authentic environment, chatbots usually offer a more limited version of such experiences. From the perspective of extended mind epistemology, disembodied empathy can be understood as an artificial extension of the human cognitive and emotional process into the digital realm.

Digital tools can enhance our cognitive systems, aiding memory, decision-making, and empathy. Modern chatbots often incorporate human-like design, adaptability, proactivity, transparency, privacy, ethics, and relationship-building (Strohmann, 2023). By simulating empathy, chatbots help users feel understood and valued, encouraging continuous engagement in therapy. Recognizing users' emotions fosters trust and persistence in the therapeutic process. Additionally, the non-judgmental nature of chatbots allows users to discuss sensitive topics more openly, enhancing therapy's effectiveness.

## 5.7 Simplicity and specificity

General-use chatbots are often developed with more extensive datasets and sophisticated algorithms. These chatbots are designed to operate across a wide range of domains, which means they benefit from diverse forms of data.

Paradoxically, despite the advanced capabilities of general-use AI chatbots for cognitive restructuring, it is feasible that many users may prefer simpler, specialized therapeutic chatbots.

Simpler bots provide straightforward interactions, making them easier for users to understand and follow, avoiding patronizing explanations. Chatbot are also reported to significantly reduce users' anticipated communication quality (Zhou et al., 2023). Their use of less complex language and concepts is beneficial for those new to digital therapeutics. They are often more suitable for users seeking emotional support rather than complex cognitive reframing. Simpler bots are more predictable and consistent, but they often provide less subtle and simplistic cognitive restructuring.

## 5.8 Limitations

The sample size of users was relatively small, with 6 virtual cases per each of the five chatbots tested. Although this provides a basic framework for comparison, a larger sample size could offer more robust and generalizable results. The study's design involved six distinct biases, each tested across five standardized prompts, which may not encompass the full spectrum of potential interactions and outcomes in real-world scenarios.

Additionally, the use of standardized prompts and specific evaluation criteria could limit the scope of the chatbot responses, potentially affecting their adaptability to varied user inputs. The evaluation process involved assessments by two cognitive scientists and a super-evaluator therapist, introducing subjective elements that might influence the results. Although these experts add credibility, inherent biases in their evaluations could skew the outcomes. Experts may also have preconceived notions about therapeutic versus non-therapeutic chatbots, leading them to favor responses that align with their expectations or professional experiences, although in this case, the obtained results contradict this possibility.

Furthermore, the study focused solely on chatbot performance and affect recognition without examining user satisfaction or real-world therapeutic impact, which are crucial metrics for gauging the practical effectiveness of chatbots.

## 6 Conclusions

The study indicates that while therapeutic chatbots hold promise in supporting mental health interventions by addressing cognitive biases, there remains a significant gap between their potential and current capabilities.

1) The interaction between AI and cognitive biases highlights the need for AI systems that not only correct but also understand and support these processes. This advocates for cautious optimism in using AI technology, emphasizing solutions that respect cognitive biases as part of the human cognitive repertoire (Rządeczka, Wodziński & Moskalewicz, 2023).

2) The high degree of bias perpetuation suggests a need for further refinement in enhancing simulated emotional intelligence and personalized response mechanisms.

3) The use of general use chatbots like GPT or Gemini for mental health feedback raises ethical concerns about boundary violations and expertise overreach. Despite disclaimers stating they are not meant for therapy; users may disregard these warnings and trust them for mental health advice. This misleads users into treating chatbots as authoritative sources, potentially worsening their issues. AI developers must ensure their tools do not become de facto mental health advisers without safeguards. Relying solely on disclaimers is insufficient; more robust measures are needed to prevent chatbots from engaging in mental health discussions, even if it means refusing to answer such questions.

4) Therapeutic chatbots aim to minimize user discomfort, but this approach can be suboptimal for effective therapy. Effective therapy often requires significant initial effort from individuals, fostering engagement and adherence. Chatbots' focus on comfort may hinder this process, leading to user overdependence on the bot for emotional support and life-coaching. This overdependence can prevent users from independently facing and overcoming their mental health challenges, which is crucial for therapeutic progress.

5) Some biases can be therapeutically beneficial, such as those related to self-esteem. Minimizing these may cause more harmful biases to emerge. Bots can be tools for continuous monitoring, providing support when a therapist is unavailable to prevent issues like self-harm. They may be beneficial for anxiety or depressive disorders but could perpetuate delusions in schizophrenia (Østergaard, 2023). Excessive or inappropriate use of chatbots may worsen mental health conditions, making it crucial to enhance affect recognition and minimize bias reinforcement in chatbot design for safe and effective use, especially in vulnerable groups.

# Author contributions

MR: Conceptualization, Formal analysis, Investigation, Methodology, Validation, Visualisation, Writing – original draft, Writing – review & editing. AS: Investigation, Methodology. PK: Investigation, Methodology. JS: Investigation, Methodology. MM: Conceptualization, Data curation, Funding acquisition, Methodology, Project administration, Supervision, Validation, Writing – review & editing.

# Funding

The author(s) declare financial support was received for the research, authorship, and/or publication of this article. The research was funded by IDEAS NCBR. Marcin Moskalewicz was supported by the Alexander von Humboldt Foundation. Marcin Rządeczka was supported by the National Science Centre, Poland, under Grant No. 2023/07/X/HS1/01557, MINIATURA 7.

# Conflict of interest

The authors declare that the research was conducted in the absence of any commercial or financial relationships that could be construed as a potential conflict of interest.

# References

1. Abd-Alrazaq, A., Alajlani, M., Ali, N., Denecke, K., Bewick, B. M., & Househ, M. (2021). Perceptions and opinions of patients about mental health chatbots: scoping review. Journal of Medical Internet Research, 23(1), e17828. https://doi.org/10.2196/17828
2. Artino, A. R., Jr, Durning, S. J., Waechter, D. M., Leary, K. L., & Gilliland, W. R. (2012). Broadening our understanding of clinical quality: from attribution error to situated cognition. Clinical pharmacology and therapeutics, 91(2), 167–169. https://doi.org/10.1038/clpt.2011.229
3. Aghakhani, S., Carre, N., Mostovoy, K., Shafer, R., Baeza-Hernandez, K., Entenberg, G., Testerman, A., & Bunge, E. L. (2023). Qualitative analysis of mental health conversational agents messages about autism spectrum disorder: a call for action. Frontiers in digital health, 5, 1251016. https://doi.org/10.3389/fdgth.2023.1251016
4. Balcombe, L. (2023). "AI Chatbots in Digital Mental Health" Informatics 10, no. 4: 82. https://doi.org/10.3390/informatics10040082
5. Beatty C, Malik T, Meheli S, Sinha C. Evaluating the therapeutic alliance with a free-text CBT conversational agent (wysa): a mixed-methods study. Front Digit Health. (2022) 4:847991. https://doi.org/10.3389/fdgth.2022.847991/full
6. Brännström, A., Wester, J., & Nieves, J. C. (2022). A formal understanding of computational empathy in interactive agents. SSRN Electronic Journal. https://doi.org/10.2139/ssrn.4138212
7. Byom, L. J., & Mutlu, B. (2013). Theory of mind: mechanisms, methods, and new directions. Frontiers in human neuroscience, 7, 413. https://doi.org/10.3389/fnhum.2013.00413
8. Cameron, G., Cameron, D. M., Megaw, G., Bond, R., Mulvenna, M., O'Neill, S., … & McTear, M. F. (2019). Assessing the usability of a chatbot for mental health care. Internet Science, 121-132. https://doi.org/10.1007/978-3-030-17705-8_11
9. Cowan, B. R., Clark, L., Candello, H., & Tsai, J. (2023). Introduction to this special issue: guiding the conversation: new theory and design perspectives for conversational user interfaces. Human–Computer Interaction, 38(3–4), 159–167. https://doi.org/10.1080/07370024.2022.2161905
10. Chan, W., Fitzsimmons-Craft, E., Smith, A., Firebaugh, M., Fowler, L., DePietro, B., … & Jacobson, N. (2022). The challenges in designing a prevention chatbot for eating disorders: observational study. Jmir Formative Research, 6(1), e28003. https://doi.org/10.2196/28003
11. D'Alfonso, S., Lederman, R., Bucci, S., Berry, K. (2020). The Digital Therapeutic Alliance and Human-Computer Interaction. JMIR Ment Health, 7(12):e21895. https://doi.org/10.2196/21895
12. Darcy A, Daniels J, Salinger D, Wicks P, Robinson A. Evidence of human-level bonds established with a digital conversational agent: cross-sectional, retrospective observational study. JMIR Form Res. (2021) 5(5):e27868. https://formative.jmir.org/2021/5/e27868
13. Damij, N. and Bhattacharya, S. D. (2022). The role of ai chatbots in mental health related public services in a (post)pandemic world: a review and future research agenda. 2022 IEEE Technology and Engineering Management Conference (TEMSCON EUROPE). https://doi.org/10.1109/temsconeurope54743.2022.9801962
14. Dosovitsky, G., Pineda, B. S., Jacobson, N. C., Chang, C., Escoredo, M., & Bunge, E. L. (2020). Artificial Intelligence Chatbot for Depression: Descriptive Study of Usage. JMIR formative research, 4(11), e17065. https://doi.org/10.2196/17065

https://doi.org/10.48550/arXiv.2406.13813

15. Durt, Ch. forthcoming 2024. "Die Digitalisierung der Lebenswelt: Von der Mathematisierung der Natur zur intelligenten Manipulation des menschlichen Sinn- und Erlebenshorizontes." In Digitale Lebenswelt. Philosophische Perspektiven. Hrsg. Maria Schwartz, Meike Neuhaus, und Samuel Ulbricht. Berlin: J.B. Metzler.
16. Fenton-O'Creevy, M., Nicholson, N., Soane, E., & Willman, P. (2003). Trading on illusions: unrealistic perceptions of control and trading performance. Journal of Occupational and Organizational Psychology, 76(1), 53-68. https://doi.org/10.1348/096317903321208880
17. Franze, A., Galanis, C. R., & King, D. L. (2023). Social chatbot use (e.g., ChatGPT) among individuals with social deficits: Risks and opportunities. Journal of Behavioral Addictions, 12(4), 871-872. https://doi.org/10.1556/2006.2023.00057
18. Fuchs, Thomas (2014). The Virtual Other: Empathy in the Age of Virtuality. Journal of Consciousness Studies 21 (5-6):152-173.
19. Gabriel et al. (2024). The Ethics of Advanced AI Assistants. https://storage.googleapis.com/deepmind-media/DeepMind.com/Blog/ethics-of-advanced-ai-assistants/the-ethics-of-advanced-ai-assistants-2024-i.pdf
20. Gamble, A. (2020). Artificial intelligence and mobile apps for mental healthcare: a social informatics perspective. Aslib Journal of Information Management, 72(4), 509-523. https://doi.org/10.1108/ajim-11-2019-0316
21. Ghassemi, M., Naumann, T., Schulam, P., Beam, A. L., Chen, I. Y., & Ranganath, R. (2020). A Review of Challenges and Opportunities in Machine Learning for Health. AMIA Joint Summits on Translational Science proceedings. AMIA Joint Summits on Translational Science, 2020, 191–200.
22. Grodniewicz, J.P., Hohol, M. Therapeutic Chatbots as Cognitive-Affective Artifacts. Topoi (2024). https://doi.org/10.1007/s11245-024-10018-x.
23. Grové, C. (2021). Co-developing a mental health and wellbeing chatbot with and for young people. Frontiers in Psychiatry, 11. https://doi.org/10.3389/fpsyt.2020.606041
24. Görnemann, E., & Spiekermann, S. (2022). Emotional responses to human values in technology: The case of conversational agents. Human–Computer Interaction, 1–28. https://doi.org/10.1080/07370024.2022.2136094
25. Habicht, J., Viswanathan, S., Carrington, B., Hauser, T. U., Harper, R., & Rollwage, M. (2024). Closing the accessibility gap to mental health treatment with a personalized self-referral chatbot. Nature medicine, 30(2), 595–602. https://doi.org/10.1038/s41591-023-02766-x
26. Harding, W. G., McConatha, J. T., & Kumar, V. K. (2020). The Relationship between Just World Beliefs and Life Satisfaction. International journal of environmental research and public health, 17(17), 6410. https://doi.org/10.3390/ijerph17176410
27. Haque, M. D. R., Rubya, S. (2023). An Overview of Chatbot-Based Mobile Mental Health Apps: Insights From App Description and User Reviews. JMIR Mhealth Uhealth 2023;11:e44838. https://doi.org/10.2196/44838
28. He, Y., Li, Y., Zhu, X., Wu, B., Zhang, S., Qian, C., … & Tian, T. (2022). Mental health chatbot for young adults with depressive symptoms during the covid-19 pandemic: single-blind, three-arm randomized controlled trial. Journal of Medical Internet Research, 24(11), e40719. https://doi.org/10.2196/40719
29. Huang S, Lai X, Ke L, Li Y, Wang H, Zhao X, Dai X, Wang Y. (2024). AI Technology panic—is AI Dependence Bad for Mental Health? A Cross-Lagged Panel Model and the Mediating Roles of Motivations for AI Use Among Adolescents. Psychol Res Behav Manag. 17:1087-1102. https://doi.org/10.2147/PRBM.S440889
30. Ismael, M., Hashim, N., Shah, N., & Munir, N. (2022). Chatbot system for mental health in bahasa malaysia. Journal of Integrated and Advanced Engineering (Jiae), 2(2), 135-146. https://doi.org/10.51662/jiae.v2i2.83

https://doi.org/10.48550/arXiv.2406.13813

31. Jain, G., Pareek, S., & Carlbring, P. (2024). Revealing the source: how awareness alters perceptions of ai and human-generated mental health responses. Internet Interventions, 36, 100745. https://doi.org/10.1016/j.invent.2024.100745
32. Khawaja, Z., & Bélisle-Pipon, J. C. (2023). Your robot therapist is not your therapist: understanding the role of AI-powered mental health chatbots. Frontiers in digital health, 5, 1278186. https://doi.org/10.3389/fdgth.2023.1278186
33. Konya-Baumbach, E., Biller, M., von Janda, S. (2023). Someone out there? A study on the social presence of anthropomorphized chatbots. Computers in Human Behavior, Volume 139, 107513. https://doi.org/10.1016/j.chb.2022.107513
34. Laakasuo, M., Herzon, V., Perander, S. et al. Socio-cognitive biases in folk AI ethics and risk discourse. AI Ethics 1, 593–610 (2021). https://doi.org/10.1007/s43681-021-00060-5
35. Leo, A. J., Schuelke, M. J., Hunt, D. M., Miller, J. P., Areán, P. A., & Cheng, A. L. (2022). Digital Mental Health Intervention Plus Usual Care Compared With Usual Care Only and Usual Care Plus In-Person Psychological Counseling for Orthopedic Patients With Symptoms of Depression or Anxiety: Cohort Study. JMIR formative research, 6(5), e36203. https://doi.org/10.2196/36203
36. Marciano, L. and Saboor, S. (2023). Reinventing mental health care in youth through mobile approaches: current status and future steps. Frontiers in Psychology, 14. https://doi.org/10.3389/fpsyg.2023.1126015
37. Moran, J., Jolly, E., & Mitchell, J. (2014). Spontaneous mentalizing predicts the fundamental attribution error. Journal of Cognitive Neuroscience, 26(3), 569-576. https://doi.org/10.1162/jocn_a_00513
38. Noble, J. M., Zamani, A., Gharaat, M., Merrick, D., Maeda, N., Lambe Foster, A., Nikolaidis, I., Goud, R., Stroulia, E., Agyapong, V. I. O., Greenshaw, A. J., Lambert, S., Gallson, D., Porter, K., Turner, D., & Zaiane, O. (2022). Developing, Implementing, and Evaluating an Artificial Intelligence-Guided Mental Health Resource Navigation Chatbot for Health Care Workers and Their Families During and Following the COVID-19 Pandemic: Protocol for a Cross-sectional Study. JMIR research protocols, 11(7), e33717. https://doi.org/10.2196/33717
39. Ogilvie, L., Prescott, J., & Carson, J. (2022). The Use of Chatbots as Supportive Agents for People Seeking Help with Substance Use Disorder: A Systematic Review. European addiction research, 28(6), 405–418. https://doi.org/10.1159/000525959
40. Østergaard, S. D. (2023). Will Generative Artificial Intelligence Chatbots Generate Delusions in Individuals Prone to Psychosis?, Schizophrenia Bulletin, Volume 49, Issue 6, 1418–1419, https://doi.org/10.1093/schbul/sbad128
41. Park, G., Chung, J., & Lee, S. (2022). Effect of ai chatbot emotional disclosure on user satisfaction and reuse intention for mental health counseling: a serial mediation model. Current Psychology, 42(32), 28663-28673. https://doi.org/10.1007/s12144-022-03932-z
42. Potts, C., Lindström, F., Bond, R., Mulvenna, M., Booth, F., Ennis, E., … & O'Neill, S. (2023). A multilingual digital mental health and well-being chatbot (chatpal): pre-post multicenter intervention study. Journal of Medical Internet Research, 25, e43051. https://doi.org/10.2196/43051
43. Rządeczka, M., Wodziński, M., & Moskalewicz, M. (2023). Cognitive biases as an adaptive strategy in autism and schizophrenia spectrum: the compensation perspective on neurodiversity. Frontiers in Psychiatry, 14. https://doi.org/10.3389/fpsyt.2023.1291854
44. Schick A, Feine J, Morana S, Maedche A, Reininghaus U (2022). Validity of Chatbot Use for Mental Health Assessment: Experimental Study. JMIR Mhealth Uhealth 10(10): e28082. https://doi.org/10.2196/28082
45. Schillings, C., Meißner, E., Erb, B., Bendig, E., Schultchen, D., Pollatos, O. (2024). Effects of a Chatbot-Based Intervention on Stress and Health-Related Parameters in a Stressed Sample: Randomized Controlled Trial. JMIR Ment Health 2024;11:e50454 https://doi.org/10.2196/50454

https://doi.org/10.48550/arXiv.2406.13813

46. Sedlakova J, Trachsel M. Conversational artificial intelligence in psychotherapy: a new therapeutic tool or agent? Am J Bioeth. (2023) 23(5):4–13. https://doi.org/10.1080/15265161.2022.2048739
47. Stubbe D. E. (2018). The Therapeutic Alliance: The Fundamental Element of Psychotherapy. Focus (American Psychiatric Publishing), 16(4), 402–403. https://doi.org/10.1176/appi.focus.20180022
48. Strohmann, T., Siemon, D., Khosrawi-Rad, B., & Robra-Bissantz, S. (2022). Toward a design theory for virtual companionship. Human–Computer Interaction, 38(3–4), 194–234. https://doi.org/10.1080/07370024.2022.2084620
49. Szalai J. (2021). The potential use of artificial intelligence in the therapy of borderline personality disorder. Journal of evaluation in clinical practice, 27(3), 491–496. https://doi.org/10.1111/jep.13530
50. Thieme, A., Hanratty, M., Lyons, M., Palacios, J., Marques, R. F., Morrison, C., & Doherty, G. (2023). Designing human-centered AI for mental health: Developing clinically relevant applications for online CBT treatment. ACM Transactions on Computer-Human Interaction, 30(2), 1–50. https://doi.org/10.1145/3564752
51. Troncoso A, Soto V, Gomila A, Martínez-Pernía D. Moving beyond the lab: investigating empathy through the Empirical 5E approach. Front Psychol. 2023 Jul 13;14:1119469. https://doi.org/10.3389/fpsyg.2023.1119469
52. Urquiza-Haas, E. G. and Kotrschal, K. (2015). The mind behind anthropomorphic thinking: attribution of mental states to other species. Animal Behaviour, 109, 167-176. https://doi.org/10.1016/j.anbehav.2015.08.011
53. Wang, L., Touré-Tillery, M. & McGill, A.L. The effect of disease anthropomorphism on compliance with health recommendations. J. of the Acad. Mark. Sci. 51, 266–285 (2023). https://doi.org/10.1007/s11747-022-00891-6
54. Weng, J., Y, H., Heaukulani, C., Tan, C., Chang, J., Phang, Y., … & Morris, R. (2023). Mental wellness self-care in singapore with mindline.sg: a framework for the development of a digital mental health platform for behaviour change (preprint).. https://doi.org/10.2196/preprints.45761
55. Yarritu, I., Matute, H., & Vadillo, M. A. (2014). Illusion of control: the role of personal involvement. Experimental psychology, 61(1), 38–47. https://doi.org/10.1027/1618-3169/a000225
56. Zhou, Q., Li, B., Han, L., Jou, M. (2023). Talking to a bot or a wall? How chatbots vs. human agents affect anticipated communication quality. Computers in Human Behavior, Volume 143, 107674. https://doi.org/10.1016/j.chb.2023.107674
57. Zhu, Y., Janssen, M., Wang, R., & Liu, Y. (2021). It is me, chatbot: working to address the covid-19 outbreak-related mental health issues in china. user experience, satisfaction, and influencing factors. International Journal of Human-Computer Interaction, 38(12), 1182-1194. https://doi.org/10.1080/10447318.2021.1988236